# Leveraging Data and Analytics for Digital Business Transformation through DataOps: An Information Processing Perspective

**Full research paper**

**Jia Xu**
School of Computing and Information Systems
University of Melbourne
Melbourne, Australia
Email: jx3@student.unimelb.edu.au

**Humza Naseer**
School of Accounting, Information Systems and Supply Chain
RMIT University
Melbourne, Australia
Email: humza.naseer@rmit.edu.au

**Sean Maynard**
School of Computing and Information Systems
University of Melbourne
Melbourne, Australia
Email: sean.maynard@unimelb.edu.au

**Justin Filippou**
School of Computing and Information Systems
University of Melbourne
Melbourne, Australia
Email: justin.filippou@unimelb.edu.au

## Abstract

Digital business transformation has become increasingly important for organizations. Since transforming business digitally is an ongoing process, it requires an integrated and disciplined approach. Data Operations (DataOps), emerging in practice, can provide organizations with such an approach to leverage data and analytics for digital business transformation. This paper proposes a framework that integrates digital business transformation, data analytics, and DataOps through the lens of information processing theory (IPT). The details of this framework explain how organizations can employ DataOps as an integrated and disciplined approach to understand their analytical information needs and develop the analytical information processing capability required for digital business transformation. DataOps-enabled digital business transformation, in turn, improves organizational performance by improving operational efficiency and creating new business models. This research extends current knowledge on digital transformation by bringing in DataOps and analytics through IPT and thereby provides organizations with a novel approach for their digital business transformations.

# 1 Introduction

Digital business transformation is becoming a strategic imperative for most organizations. According to the Gartner Survey on organizational digital business initiatives, 69% of boards of directors have recently decreed digital acceleration and 48% of respondents have increased investment in the digital business (Lyengar et al. 2021). Organizations invest in digital technologies such as the Internet of Things, cloud, big data analytics, and mobile devices to transform their business and thereby improve customers' experience, achieve internal operational efficiency, deliver new products or services, and innovate with new business models (Fitzgerald et al. 2013; Westerman et al. 2014). Although emerging technologies bring benefits for organizations, they also cause uncertainties such as unexpected competition from new entrants (Warner and Wäger 2019). Facing opportunities but also uncertainties, organizations should adjust their business to survive in the dynamic digital environment (Sia et al. 2016).

Digital business transformation refers to how organizations can alter their value creation and change the scope of their business in this digital age (Hess et al. 2016). The ability to manage and exploit data is becoming a competitive competency in digital business transformation for organizations (Dremel et al. 2017). From this perspective, data analytics, as one of the disruptive technologies in the digital economy, is key to digital business transformation (Pappas et al. 2018). It can be defined as the extensive use of data, statistical and quantitative analysis, explanatory and predictive models (Davenport and Harris 2017) to unpack opportunities and address uncertainties for digital business transformation. Analytics enables organizations to get useful information from data for digital business transformation. For instance, organizations can get insights derived from customer interactions over digital channels to transform sales processes more effectively (Dremel et al. 2017). For organizations facing a high volume and wide variety of data generated every day, a strong analytics capability consisting of relevant people, processes, technologies, and organizations (Vidgen et al. 2017) helps to improve the capacity to process the data for transformation.

Drawing on information processing theory (IPT), this study aims to explore how organizations leverage data and analytics for digital business transformation. In IPT, uncertainty refers to the absence of information (Daft and Lengel 1986). Organizations need to process a large amount of information when coping with tasks in an uncertain environment (Galbraith 1974). In the context of digital business transformation, organizations may not have enough information to support the transformation of their business due to the emerging disruptive technologies, changing diffusion of technologies and customers' expectations (Ismail et al. 2017; Matt et al. 2015). To address these uncertainties in digital business transformation, extensive useful information from analytics is needed. For example, organizations may need analytical information about their current operations assessment to identify bottlenecks that can be transformed by digital technologies (Sia et al. 2016). Analytical information about customers' demand is also needed to sense any opportunity for new business models (Loebbecke and Picot 2015). According to IPT, to achieve a given level of performance, organizations should strive for the fit between information processing needs and information processing capability (Flynn and Flynn 1999; Galbraith 1974). This means that organizations should develop their analytical information processing capabilities that can fulfil their analytical information needs for their digital business transformation.

However, although data and analytics provide great potential for digital business transformation, there are still issues (e.g., poor data quality and lack of collaboration between stakeholders) (Ereth and Eckerson 2018). These issues would negatively impact the quality of analytical information and the analytical information processing capability, which subsequently inhibits organizations from leveraging data and analytics for digital business transformation. Therefore, data operations (DataOps) is needed. DataOps is an emerging method for analytical solutions delivery (Ereth and Eckerson 2018; Heudecker et al. 2020). By borrowing the principles and practices from methods (e.g., DevOps and Agile) used in the software engineering and manufacturing industry, DataOps aims to better manage data, provide high-quality insights, and thus improve the quality and speed of analytics for digital business transformation in a dynamic environment (Ereth 2018; Heudecker et al. 2020). Thus, DataOps has the potential to help organizations realize the benefits of analytics by getting useful analytical information and developing analytical information processing capability for digital business transformation.

The existing literature on digital business transformation mainly focuses on the strategy of digital transformation (Hess et al. 2016; Ismail et al. 2017) and the impact of digital technologies on the organization, industry, and society (Nambisan et al. 2019). Since transforming business digitally is an ongoing and long-term process (Davenport and Westerman 2018), it requires an integrated approach to help organizations leverage data and analytics for digital business transformation. However, there is very little research that explains how organizations utilize data and analytics to understand their analytical information needs and develop the analytical information processing capability required for digital business transformation. Driven by the research gap and the importance of digital business transformation, this research proposes the following research question: *How can organizations leverage data and analytics for digital business transformation through DataOps*? To address the research question, we develop a framework that integrates digital business transformation, analytics, and DataOps through the lens of IPT. The details of the framework explain how organizations can employ DataOps as a disciplined approach to understand their analytical information needs and develop the analytical information processing capability required for digital business transformation. DataOps-enabled digital business transformation, in turn, improves organizational performance by improving operational efficiency, creating new business models, and getting innovative.

This paper proceeds as follows: in the next sections, we present the method of literature review. We then give the background of this research, followed by an explanation of the framework. Finally, we conclude the paper by providing contributions and limitations of our work.

## 2　Research Method

A literature review was conducted to understand current knowledge on digital business transformation, data analytics, and DataOps following the procedures outlined by Webster and Watson's (2002) methodology. As a first step, we searched literature from popular literature databases (e.g., Science Direct and AIS Electronic library) using the combination of search terms such as ("digital transformation" AND "analytics") and ("analytics" AND "DataOps"). These searches identified a total of 200 articles. We went through the titles and abstracts of each article to examine if it was relevant to the research question (i.e., how organizations can leverage data and analytics for digital business transformation through DataOps). We identified 39 articles on the topics of digital transformation and analytics. However, as DataOps is an emerging concept, there was little academic literature on DataOps (11 articles) relevant to our research question. To expand further, we also reviewed 10 relevant articles from practitioner research and advisory firms, such as Gartner, DataKitchen and The Eckerson Group, to understand how DataOps plays a role in digital business transformation. After the initial selection, there were 60 articles for review and coding. Following the guidelines from Webster and Watson (2002), we went backward by reviewing the reference list in the article identified in the previous step to include another 10 articles we should consider. The whole process resulted in 70 articles for in-depth review and coding. A concept-centric approach from Watson and Webster (2020) was used to categorize literature. We synthesized, analysed and integrated the literature using information processing theory and classified it into the paths shown in Figure 1. Out of 70 coded articles, 51 included variables of interest and are compiled in the analysis.

## 3　Background

### 3.1　Employing DataOps for Data and Analytics

A high volume and wide variety of data is generated in the current digital age, causing data and analytics to gain increasing attention from organizations (Pappas et al. 2018). However, processing big data and getting insights from analytics are both challenging. Wells (2019) lists the challenges - including rapidly increasing data volumes, more sources of data, more data use cases, and more stakeholders and data consumers in modern data ecosystems. These challenges will increase risks of conflicting and wrong data, data silos, and a long waiting time to get the needed information, which subsequently inhibits organizations extracting value from data and analytics. DataOps, therefore, emerges in practice with the aim to help organizations overcome challenges in modern analytical ecosystems and thus benefit from data and analytics.

DataOps was popularized by Palmer (2015), who explains that DataOps acknowledges the interconnected nature of data engineering, integration, quality, security and privacy to help organizations accelerate analytics and enable previously impossible analytics tasks. DataOps applies the best principles and practices from methods (i.e., Agile, DevOps, Lean, and Total Quality Management) used in the software engineering and manufacturing areas to data analytics. It enables organizations to take the pain out of data analytics, streamline the process of analytics, and maximize the business value of data as well as improve satisfaction (Eckerson 2019). As DataOps is a rather novel concept, there is no generally accepted definition (Ereth 2018). However, since DataOps was coined, it has been most closely associated with the quality and efficiency in the delivery of analytics (Aslett 2020). From the perspective of quality, DataOps helps to promote communication between, and integration of, formerly siloed data, teams, and systems (Thusoo and Sarma 2017), which reduces the risks of conflicting and wrong results from analytics and thus ensures the quality of analytics. From the perspective of efficiency,

DataOps represents a culture change that focuses on improving collaboration and accelerating analytical delivery by adopting lean or iterative practices where appropriate (Heudecker et al. 2020). Therefore, in our research, DataOps is defined as an integrated approach for delivering data analytical solutions and building analytics capabilities in a way that continuously accelerates output and improves the quality of analytics (Ereth and Eckerson 2018; Naseer et al. 2020).

## 3.2　The Role of Data and Analytics in Digital Business Transformation

The concept of digital business transformation has evolved from digitization to digitalization and finally, digital transformation (Legner et al. 2017). These concepts are closely associated and even sometimes incorrectly used interchangeably. Therefore, it is important to differentiate between these concepts. Digitization refers to the encoding of analogue information into digital format (Yoo 2010). Digitalization is about how IT or digital technologies can be used to alter existing business processes for not only cost savings but also process improvements (Verhoef et al. 2021). Compared with digitization and digitalization, digital transformation may have broader organizational implications. It refers to a company-wide phenomenon with broad organizational implications in which, most notably, the core business model of the firm is subject to change by using digital technologies (Hess et al. 2016; Verhoef et al. 2021). Digitization, digitalization, and digital transformation can be viewed as three phases of the transformation journey, in which organizations "may start with minor changes (e.g., digitization or digitalization) to gradually transform their traditional business into a digital one" (Verhoef et al. 2021, p. 892). Our research focuses on how organizations transform their business in the digital era. Therefore, we use the term of digital business transformation to differentiate it from digitization and digitalization.

As mentioned before, in digital business transformation, an organization's business model is likely to change. Such changes can be done through undertaking experiments for a new business model or improving the current business model (Warner and Wäger 2019). Incremental enhancements of current business models aim at optimizing existing processes to increase overall efficiency and quality of products and services, which may replace less efficient business models in the long run (Loebbecke and Picot 2015). Digitization and digitalization have fostered the generation of big data, which emphasizes the volume, velocity, variety, veracity, variability, and value of data (Conboy et al. 2020; Seddon and Currie 2017). Analytics, which refers to the extensive use of data, statistical and quantitative analysis, explanatory and predictive models (Davenport and Harris 2017), offers organizations opportunities to utilize the big data for their digital business transformations (e.g., improving the current business models or creating new business models). Table 1 lists examples of how data and analytics can be used for digital business transformation.

| Reference | Industry | Data | Analytics | Digital Business Transformation |
|---|---|---|---|---|
| Frank et al. (2019) | Manufacturing | Product-related data from embedded sensors | Use analytical algorithms to extract patterns of product usage | Segment market and identify opportunities for new product development and new services |
| Conboy et al. (2020) | Telecommunications | Data generated by interactions with users on social media | Use social media analytics to detect best practices and methods for marketing approaches | Improve customers satisfaction and position itself as the largest service provider in the market |
| Müller et al. (2016) | Technology provider | Customers' service requests and feedback | Use text analytics to monitor the feedback of thousands of customers in real-time and respond to patterns and trends | Achieve efficiency gains and improve customer service |

*Table 1. Concept Matrix: Using Data and Analytics for Digital Business Transformation*

However, digital business transformation is a long-term process and journey. As emphasized by Warner and Wäger (2019, p. 327), "the genuine digital transformations are an ongoing process of using digital technologies in everyday organizational life". The long-term game of digital business transformation requires not only extensive investment and time but also a holistic and integrated approach to make steady progress toward the right end state (Davenport and Westerman 2018; Sia et al. 2016). In this research, we argue that DataOps can act as such an integrated approach, which enables organizations to unpack the value of data and leverage analytics for digital business transformation.

## 3.3 Conceptualizing Digital Business Transformation through Information Processing Theory

Organizations can be viewed as an information processing system facing complexity and uncertainty (Tushman and Nadler 1978). According to Galbraith (1974, p. 28), "the greater the uncertainty is, the greater the amount of information is needed to be processed among decision-makers to achieve a given level of performance". To get the needed information, organizations need to develop their information processing capabilities: the ability to gather, transform, interpret, synthesize, analyse, store, and communicate data, information, and knowledge to cope with the variety, uncertainty and an unclear environment (Naseer et al. 2021; Tushman and Nadler 1978). Effective organizations are those that fit their information processing capabilities to their information needs, which means that organizations must design and adopt appropriate structures, mechanisms, and practices to develop the information processing capabilities that meet and satisfy the information processing needs (Flynn and Flynn 1999; Galbraith 1974; Tushman and Nadler 1978).

In the context of digital business transformation, when organizations begin their digital business transformation journeys, they are very likely to encounter a high degree of uncertainty and lack of enough information to cope with uncertainty in the dynamic environment (Matt et al. 2015; Vial 2019). Therefore, to reduce uncertainty, relevant information and insights such as the current business performance and potential revenue stream are needed to make informed decisions so that organizations can better transform their business (Ismail et al. 2017). Moreover, according to Li et al. (2021), the information processing capability enabled by digital technologies can positively impact business transformation by enhancing market agility. Given the potential of analytics and DataOps, organizations can leverage data and analytics through DataOps to understand their analytical information needs and develop their analytical information processing capability required for digital business transformation. With the needed analytical information and required capability to gather, transform, store, and analyse data for digital business transformation, organizations are more likely to optimize their digital business transformation.

## 4 Digital Business Transformation through DataOps: A Theoretical Framework

Based on our synthesis, analysis, and integration of the literature, we develop a theoretical framework that links digital business transformation, analytics, and DataOps through the lens of IPT (see Figure 1). In this framework, the dynamic digital environment triggers the digital business transformation journey of organizations. To cope with uncertainties in digital business transformation, organizations need to understand their analytical information needs and develop an analytical information processing capability to fulfil the needs by utilising data and analytics. DataOps provides organizations with an integrated and disciplined approach to developing their analytical information processing capabilities for digital business transformation. DataOps-enabled digital business transformation can then help to improve organizational performance (e.g., creating new business models). Thus, DataOps indirectly impacts organizational performance through the mediation of digital business transformation. Table 2 gives definitions of the concepts in this framework.

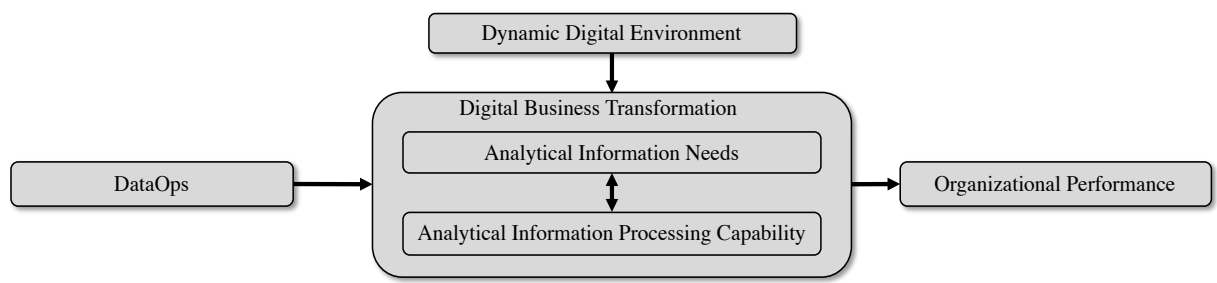

*Figure 1: A Framework for Digital Business Transformation through DataOps*

| Concept | Definition | Reference |
|---|---|---|
| Dynamic Digital Environment | A dynamic environment consisting of disruptive digital technologies, new competitors, and changing customers' expectations. | (Vial 2019; Warner and Wäger 2019) |
| DataOps | An integrated and disciplined approach to developing analytical information processing capability for digital business transformation | (Naseer et al. 2020) |
| Analytical Information Needs | The analytical information that can help to reduce the uncertainties and unpack opportunities in digital business transformation | (Ashrafi et al. 2019; Dremel et al. 2017) |
| Analytical Information Processing Capability | The ability to use analytical practices and applications to capture, integrate, analyse, and visualise data, and utilise insights for digital business transformation. | (Cao et al. 2019; Saldanha et al. 2017) |
| Organizational Performance | The positive outcome (e. g. operational efficiency, new business models, and innovativeness) brought by the digitally transformed business. | (Vial 2019) |

*Table 2. Definitions of Key Concepts in the Proposed Theoretical Framework*

## 4.1　Dynamic Digital Environment

Organizations operate in a dynamic digital environment with opportunities and uncertainties. On the one hand, the digital environment consisting of rapidly evolved digital technologies offers opportunities to transform their business. For example, Artificial Intelligence can help organizations engage customers and employees and deliver new products and services (Borges et al. 2021). Also, 3D printing can be used to produce customized products and manufacture different goods using the same resources (Frank et al. 2019). On the other hand, the advancement and disruption of novel digital technologies would intensify the environmental dynamism and thereby increase uncertainty, characterized by the entrance of new competitors and changing expectations of customers (Warner and Wäger 2019).

From the perspective of market competition, new technologies impact the existing market by creating a new market, in which the entrants would always win (Christensen 2013). Facing new competitors, organizations need to start their digital business transformation journeys to compete in the market. From the perspective of changing customers' expectations, disruptive digital technologies such as smartphones also play decisive roles in shaping and mediating all dimensions of people's lived experiences (Yoo 2010), creating change in customers' expectations and requirements (Schallmo et al. 2017; Vial 2019). Organizations need to quickly understand and integrate customers' requirements by transforming their existing services or products (Hess et al. 2016; Sia et al. 2016). Thus, we propose the following:

Proposition 1: The digital environment in which organizations operate acts as a trigger for digital business transformation.

## 4.2　Employing DataOps for Digital Business Transformation

As explained in Section 3, organizations can leverage data and analytics for digital business transformation. However, unpacking the value of data and analytics requires an integrated approach. DataOps can act as such an approach to help organizations unite stakeholders, govern the flow of data, and ensure that the insights from analytics can satisfy the needs of digital business transformation in time. Similar to DevOps which helps to build a collaboration between software development and IT operations (Munappy et al. 2020), DataOps unites data stakeholders (e. g. data engineers, data analysts, IT operations, and business users) around business requirements for digital business transformation (Ereth and Eckerson 2018). Borrowing from Total Quality Management, using statistical measurement to monitor and control manufacturing processes, DataOps verifies that results at each intermediate step in the production of analytics matches business requirements (Bergh et al. 2019). Lean thinking also makes it possible to identify waste and manage the flow of data to deliver analytical solutions as quickly as possible (Atwal 2020). Moreover, DataOps can discover common ground where various stakeholders and technologies can act in concert (e.g., orchestration tools enable automatic combinations of various technologies) (Ereth and Eckerson 2018). Last but not least, by adopting Agile and welcoming changing requirements, DataOps enables a culture of encouraging changes to help organizations cope with the dynamic digital environment when using analytics (Atwal 2020). Overall, learning from these mature methods from software engineering and manufacturing, DataOps is not only a method for analytical products development but also a method that integrates people, process, technology, and culture in analytics for digital business transformation. Therefore, we propose:

Proposition 2: DataOps provides an integrated and disciplined approach for organizations to harness data and analytics for digital business transformation.

In a dynamic digital environment, organizations may lack information for digital business transformation. For example, since the diffusion of digital technologies can change swiftly, organizations may not have enough information to make solid assumptions in organizational digital business transformation strategies (Matt et al. 2015). Further, changing customers' expectations and requirements make it hard to get sufficient information to predict how the market will change, especially when more and more born-digital firms are entering the market (Ismail et al. 2017). Moreover, there are lots of complexities in internal operations, such as the tension between exploiting the existing business while also exploring new digital initiatives that are compatible with the pact dependencies of the past (Warner and Wäger 2019). Organizations may not have adequate information to cope with complexities in digital business transformation.

From the view of IPT, facing uncertainties in digital business transformation, extensive useful information (e.g., the market trend, customers' expectations, the performance of business operations, and stakeholders' sentiments on the cultural shift towards digital business transformation) is required (Ismail et al. 2017; Loebbecke and Picot 2015). Insights drawn from analytics are likely to significantly improve the amount and richness of information for digital business transformation (Ashrafi et al. 2019; Dremel et al. 2017). For example, organizations can get insights about new customer-centric trends that are hard for strategic planners to predict from analytics (Warner and Wäger 2019). Conboy et al. (2020) also found that organizations can extract rich information (e.g., the sentiment of customers as well as staff and predicted defect rates) from analytics to understand customers and competitors and identify internal inefficiencies that need to be transformed. According to the 'fit' concept in IPT (Daft and Lengel 1986; Flynn and Flynn 1999), organizations should develop their analytical information processing capabilities to match their information needs for digital business transformation. The analytical information processing capability in our framework refers to the ability to use analytical practices and applications to process (i.e., capture, store, transform, analyse, and visualise) data, and utilise insights for digital business transformation (Cao et al. 2019; Saldanha et al. 2017). Organizations can leverage tools, techniques, people and processes in analytics to process data and thereby produce the needed insights for their business transformations (Srinivasan and Swink 2018).

The following principles and practices from DataOps play an important role in enabling organizations to get the needed analytical information and develop the analytical information processing capability required for digital business transformation:

- Business value: DataOps highlights the mindset that data is not an end in itself but should deliver insights that add value to the business (Ereth and Eckerson 2018). This mindset ensures that every stakeholder involved views data quality as a top priority to deliver analytical insights that satisfy the needs of digital business information.

- Testing and monitoring: DataOps emphasizes the importance of testing and monitoring at every stage in data processing to ensure that the analytical information delivered can satisfy the needs of digital business transformation. Testing enables quick identification of issues before they are delivered to the business and become hard to fix (Ereth and Eckerson 2018). Monitoring the performance of data movement processes is also important as it can detect unexpected patterns and problems (Ereth and Eckerson 2018).

- Continuous improvement: One of the keys in DataOps is to continuously improve (Ereth and Eckerson 2018). It requires teams to learn from mistakes, review processes continuously to adapt to the changing environment (Heudecker et al. 2020). Such an iterative process with incremental improvements helps to ensure that organizations can extract high-quality analytical information for digital business transformation in a sustainable manner.

- Automation: DataOps uses technologies to automate wherever possible (Ereth and Eckerson 2018). Automation can improve the reliability of analytical information processing capability and reduce the amount of time required for data processing, that previously relied on human intervention. The automated analytical information delivery with appropriate levels of governance and metadata also helps to improve the use and value of analytical information in a dynamic environment (Heudecker et al. 2020)

- Collaboration and communication: DataOps encourages sharing knowledge, simplifying communication, and providing feedback at every stage of analytical information processing (Ereth and Eckerson 2018, p. 8). Close collaboration and effective communication between stakeholders helps to clarify ambiguity and ensures all stakeholders have the same understanding and goals, thereby improving the efficiency of analytical information processing capabilities.

- Cultural shift: DataOps catalyses organizational and cultural shifts (Wells 2019). DataOps emphasizes that processing data is not only about technologies. Instead, all the resources including people, technologies, and processes need to be combined and orchestrated (Wells 2019) for data processing capabilities and thus deliver the needed analytical information for digital business transformation.

Accordingly, we argue:

Proposition 3: DataOps principles and practices enable organizations to improve the fit between their analytical information needs and the analytical information processing capabilities required for digital business transformation.

## 4.3 Organizational Performance

Digitally transformed business can help to improve organizational performance such as operational efficiency, new business models, and innovativeness (Vial 2019; Warner and Wäger 2019; Westerman et al. 2014). In the view of IPT, the fit between information needs and information processing capability can lead to optimal performance (Tushman and Nadler 1978). In the context of digital business transformation, achieving the fit between analytical information needs and the analytical information processing capability means that organizations can get insights from analytics that are not only useful for the transformation but also can benefit the business. For example, Conboy et al. (2020) give an example of a bank that creates a new business model and services by utilising insights to provide consultancy services to third parties. Moreover, insights from data analytics can help organizations gather information about the environment and thus develop a better understanding for innovation (Duan et al. 2020).

Moreover, organizations can utilize the analytical information processing capability to process all the data generated in digital activities, which also brings value to organizations. For example, the ability to automatically process data and make decisions using analytics helps to improve the operational efficiency for organizations (Loebbecke and Picot 2015). Overall, if organizations can transform their business so that all the analytical information needs and required analytical information processing capability for digital business transformation can be satisfied, organizations are more likely to transform their business digitally, thereby improving the organizational performance. Therefore, we propose:

Proposition 4: Digital business transformation has a positive impact on the overall organizational performance by creating new business models, improving innovation and operational efficiency.

## 4.4 Linking DataOps with Organizational Performance

According to a survey about the importance of DataOps conducted by Aslett (2020), more than 81% of respondents agreed that DataOps would have a positive impact on their organization's success. In our framework, DataOps is also indirectly linked to organizational performance. First, DataOps can help organizations better utilize data and analytics for digital business transformation. By highlighting business value, testing and monitoring, and continuous improvement, automation, collaboration and communication, and a cultural shift, DataOps helps to reduce the time to identify insights for digital business transformation, improve quality of analytical information, and orchestrate the people, process and technology for analytical information processing capabilities (Bergh et al. 2019). Then, supported by the high-quality analytical information and more reliable analytical information processing capability from DataOps, organizations can better transform their business digitally, which subsequently improves organizational performance. Therefore, we formulate the following proposition:

Proposition 5: DataOps indirectly impacts organizational performance through the mediating role of digital business transformation.

# 5 Conclusion, Limitations and Future Research

An increasing number of firms are embarking on digital business transformation journeys. However, transforming business digitally and getting value from it are challenging (Davenport and Westerman 2018). In this study, we develop a theoretical framework that explains how organizations can transform their business by leveraging data and analytics through DataOps. This framework for digital business transformation argues that dynamic digital environment triggers organizational need to obtain analytical information and develop analytical information processing capability for digital business transformation. DataOps provides an integrated and disciplined approach for organizations to harness data and analytics for digital business transformation. Principles and practices from DataOps help to improve the fit between analytical information needs and analytical information capability. Then, the digitally transformed business enabled by DataOps can help to improve organizational performance.

Our research extends the current knowledge on digital transformation by bringing in data analytics and DataOps for digital business transformation through the lens of information processing theory. Specifically, we propose DataOps as a novel approach for digital business transformation by helping organizations leverage data and analytics to get the needed analytical information and develop the analytical information processing capability. We also explain how digital business transformation can help to improve organizational performance. This study contributes to research on DataOps by linking DataOps with organizational performance through digital business transformation, which indicates the business value of DataOps for organizations. For practice, this research gives organizations an overview of how to leverage the numerous data generated in the digital world for their digital business transformations. However, this research has a few limitations that offer opportunities for future research. First, the concepts in our framework are relatively novel compared with existing literature. We expect future research to refine and evolve the concepts in this framework. Our framework does not include details on the implementation of DataOps for digital business transformation. Future research can explore the technologies, processes, and people needed to implement DataOps for digital business transformation. Moreover, future research can also conduct case studies and expert interviews to further examine and validate the framework in this research.

# 6   References

Ashrafi, A., Zare Ravasan, A., Trkman, P., and Afshari, S. 2019. "The Role of Business Analytics Capabilities in Bolstering Firms' Agility and Performance," *International Journal of Information Management* (47), pp. 1-15.

Aslett, M. 2020. "Dataops Unlocks the Value of Data," 451 Research, pp. 1-15.

Atwal, H. 2020. *Practical Dataops: Delivering Agile Data Science at Scale*. Apress.

Bergh, C., Benghiat, G., and Strod, E. 2019. *The Dataops Cookbook: Methodologies and Tools That Reduce Analytics Cycle Time While Improving Quality*. DataKitchen Headquarters.

Borges, A. F. S., Laurindo, F. J. B., Spínola, M. M., Gonçalves, R. F., and Mattos, C. A. 2021. "The Strategic Use of Artificial Intelligence in the Digital Era: Systematic Literature Review and Future Research Directions," *International Journal of Information Management* (57), pp. 1-16.

Cao, G., Duan, Y., and Cadden, T. 2019. "The Link between Information Processing Capability and Competitive Advantage Mediated through Decision-Making Effectiveness," *International Journal of Information Management* (44), pp. 121-131.

Christensen, C. M. 2013. "How Can Great Firms Fail? Insights from the Hard Disk Drive Industry," in *The Innovator's Dilemma: When New Technologies Cause Great Firms to Fail*. Harvard Business Review Press.

Conboy, K., Mikalef, P., Dennehy, D., and Krogstie, J. 2020. "Using Business Analytics to Enhance Dynamic Capabilities in Operations Research: A Case Analysis and Research Agenda," *European Journal of Operational Research* (281:3), pp. 656-672.

Daft, R. L., and Lengel, R. H. 1986. "Organizational Information Requirements, Media Richness and Structural Design," *Management science* (32:5), pp. 554-571.

Davenport, T. H., and Harris, J. 2017. *Competing on Analytics: Updated, with a New Introduction: The New Science of Winning*. Harvard Business Press.

Davenport, T. H., and Westerman, G. 2018. "Why So Many High-Profile Digital Transformations Fail," *Harvard Business Review* (9), pp. 1-5.

Dremel, C., Wulf, J., Herterich, M. M., Waizmann, J. C., and Brenner, W., 16(2). 2017. "How Audi Ag Established Big Data Analytics in Its Digital Transformation," *MIS Quarterly Executive* (16:2), pp. 81-100.

Duan, Y., Cao, G., and Edwards, J. S. 2020. "Understanding the Impact of Business Analytics on Innovation," *European Journal of Operational Research* (281:3), pp. 673-686.

Eckerson, W. 2019. "Trends in Dataops: Bringing Scale and Rigor to Data and Analytics," Eckerson Group, pp. 1-20.

Ereth, J. 2018. "Dataops-Towards a Definition," in *LWDA*, pp. 104-112.

Ereth, J., and Eckerson, W. 2018. "Dataops: Industrializing Data and Analytics Strategies for Streamlining the Delivery of Insights," Eckerson Group, pp. 1-21.

Fitzgerald, M., Kruschwitz, N., Bonnet, D., and Welch, M. 2013. "Embracing Digital Technology: A New Strategic Imperative," *MIT Sloan Management Review* (55:2), pp. 1-16.

Flynn, B. B., and Flynn, E. J. 1999. "Information‐Processing Alternatives for Coping with Manufacturing Environment Complexity," *Decision Sciences* (30:4), pp. 1021-1052.

Frank, A. G., Dalenogare, L. S., and Ayala, N. F. 2019. "Industry 4.0 Technologies: Implementation Patterns in Manufacturing Companies," *International Journal of Production Economics* (210), pp. 15-26.

Galbraith, J. R. 1974. "Organization Design: An Information Processing View," *Interfaces* (4:3), pp. 28-36.

Hess, T., Matt, C., Benlian, A., Wiesböck, F., and 15(2). 2016. "Options for Formulating a Digital Transformation Strategy," *MIS Quarterly Executive* (15:2), pp. 123-139.

Heudecker, N., Friedman, T., and Dayley, A. 2020. "Innovation Insight for Dataops," Gartner Research, pp. 1-9.

Ismail, M. H., Khater, M., and Zaki, M. 2017. "Digital Business Transformation and Strategy: What Do We Know So Far," *Cambridge Service Alliance* (10), pp. 1-36.

Legner, C., Eymann, T., Hess, T., Matt, C., Böhmann, T., Drews, P., Mädche, A., Urbach, N., and Ahlemann, F. 2017. "Digitalization: Opportunity and Challenge for the Business and Information Systems Engineering Community," *Business & Information Systems Engineering* (59:4), pp. 301-308.

Li, H., Wu, Y., Cao, D., and Wang, Y. 2021. "Organizational Mindfulness Towards Digital Transformation as a Prerequisite of Information Processing Capability to Achieve Market Agility," *Journal of Business Research* (122), pp. 700-712.

Loebbecke, C., and Picot, A. 2015. "Reflections on Societal and Business Model Transformation Arising from Digitization and Big Data Analytics: A Research Agenda," *The Journal of Strategic Information Systems* (24:3), pp. 149-157.

Lyengar, P., Furlonger, D., and Lopez, J. 2021. "Survey Analysis: Executive Leaders Should Align to Board Priorities for 2021," Gartner Research, pp. 1-25.

Matt, C., Hess, T., and Benlian, A. 2015. "Digital Transformation Strategies," *Business & Information Systems Engineering* (57:5), pp. 339-343.

Müller, O., Junglas, I., Debortoli, S., and vom Brocke, J. 2016. "Using Text Analytics to Derive Customer Service Management Benefits from Unstructured Data," *MIS Quarterly Executive* (15:4), pp. 243-258.

Munappy, A. R., Mattos, D. I., Bosch, J., Olsson, H. H., and Dakkak, A. 2020. "From Ad-Hoc Data Analytics to Dataops," in *Proceedings of the International Conference on Software and System Processes*, pp. 165-174.

Nambisan, S., Wright, M., and Feldman, M. 2019. "The Digital Transformation of Innovation and Entrepreneurship: Progress, Challenges and Key Themes," *Research Policy* (48:8), pp. 1-9.

Naseer, H., Maynard, S. B., and Desouza, K. C. 2021. "Demystifying Analytical Information Processing Capability: The Case of Cybersecurity Incident Response," *Decision Support Systems* (143), pp. 1-11.

Naseer, H., Maynard, S. B., and Xu, J. 2020. "Modernizing Business Analytics Capability with Dataops: A Decision-Making Agility Perspective," in *ECIS 2020*, pp. 1-11.

Palmer, A. 2015. "From Devops to Dataops," in *Getting Data Right: Tackling the Challenges of Big Data Volume and Variety,* C. Shannon (ed.). O'Reilly, pp. 49-57.

Pappas, I. O., Mikalef, P., Giannakos, M. N., Krogstie, J., and Lekakos, G. 2018. "Big Data and Business Analytics Ecosystems: Paving the Way Towards Digital Transformation and Sustainable Societies," *Information Systems and e-Business Management* (16:3), pp. 479-491.

Saldanha, T. J. V. V., Mithas, S., and Krishnan, M. S. 2017. "Leveraging Customer Involvement for Fueling Innovation: The Role of Relational and Analytical Information Processing Capabilities," *MIS Quarterly* (41:1), pp. 267-286.

Schallmo, D., Williams, C. A., and Boardman, L. 2017. "Digital Transformation of Business Models — Best Practice, Enablers, and Roadmap," *International Journal of Innovation Management* (21:08), pp. 1-17.

Seddon, J. J. J. M., and Currie, W. L. 2017. "A Model for Unpacking Big Data Analytics in High-Frequency Trading," *Journal of Business Research* (70), pp. 300-307.

Sia, S. K., Soh, C., and Weill, P. 2016. "How Dbs Bank Pursued a Digital Business Strategy," *MIS Quarterly Executive* (15:2), pp. 105-121.

Srinivasan, R., and Swink, M. 2018. "An Investigation of Visibility and Flexibility as Complements to Supply Chain Analytics: An Organizational Information Processing Theory Perspective," *Production and Operations Management* (27:10), pp. 1849-1867.

Thusoo, A., and Sarma, J. 2017. *Creating a Data-Driven Enterprise with Dataops.* O'Reilly Media Inc.

Tushman, M. L., and Nadler, D. A. 1978. "Information Processing as an Integrating Concept in Organizational Design.," *Academy of Management Review* (3:3), pp. 613-624.

Verhoef, P. C., Broekhuizen, T., Bart, Y., Bhattacharya, A., Qi Dong, J., Fabian, N., and Haenlein, M. 2021. "Digital Transformation: A Multidisciplinary Reflection and Research Agenda," *Journal of Business Research* (122), pp. 889-901.

Vial, G. 2019. "Understanding Digital Transformation: A Review and a Research Agenda," *The Journal of Strategic Information Systems* (28:2), pp. 118-144.

Vidgen, R., Shaw, S., and Grant, D. B. 2017. "Management Challenges in Creating Value from Business Analytics," *European Journal of Operational Research* (261:2), pp. 626-639.

Warner, K. S. R., and Wäger, M. 2019. "Building Dynamic Capabilities for Digital Transformation: An Ongoing Process of Strategic Renewal," *Long Range Planning* (52:3), pp. 326-349.

Watson, R. T., and Webster, J. 2020. "Analysing the Past to Prepare for the Future: Writing a Literature Review a Roadmap for Release 2.0," *Journal of Decision Systems* (29:3), pp. 129-147.

Webster, J., and Watson, R. T. 2002. "Analyzing the Past to Prepare for the Future: Writing a Literature Review.," *MIS Quarterly* (26:2), pp. xiii-xxiii.

Wells, D. 2019. "Intelligent Data Operations: The Next Wave in Smart Data Ecosystems," Eckerson Group, pp. 1-21.

Westerman, G., Bonnet, D., and McAfee, A. 2014. "The Nine Elements of Digital Transformation," *MIT Sloan Management Review* (55:3), pp. 1-6.

Yoo, Y. 2010. "Computing in Everyday Life: A Call for Research on Experiential Computing," *MIS Quarterly* (34:2), pp. 213-231.

## Acknowledgements

We would like to thank the University of Melbourne for providing Melbourne Research Scholarship to the first author.